\documentclass[prd,twocolumn,aps,showpacs,preprintnumbers,amsmath,amssymb]{revtex4}
\usepackage{graphicx}
\usepackage{dcolumn}
\usepackage{bm}
\usepackage{psfrag}

\begin{document}
\newcommand{\be}{\begin{equation}}\newcommand{\ee}{\end{equation}}
\newcommand{\bea}{\begin{eqnarray}}\newcommand{\eea}{\end{eqnarray}}
\newcommand{\bc}{\begin{center}}\newcommand{\ec}{\end{center}}
\def\no{\nonumber}
\def\eq#1{Eq. (\ref{#1})}\def\eqeq#1#2{Eqs. (\ref{#1}) and  (\ref{#2})}
\def\lsim{\raise0.3ex\hbox{$\;<$\kern-0.75em\raise-1.1ex\hbox{$\sim\;$}}}
\def\gsim{\raise0.3ex\hbox{$\;>$\kern-0.75em\raise-1.1ex\hbox{$\sim\;$}}}
\def\slash#1{\ooalign{\hfil/\hfil\crcr$#1$}}
\def\eff{\mbox{\tiny{eff}}}
\def\order#1{{\mathcal{O}}(#1)}
\def\pppm{B^0\to\pi^+\pi^-}
\def\pzpz{B^0\to\pi^0\pi^0}
\def\pppz{B^0\to\pi^+\pi^0}
\preprint{UCL-IPT-05-18, CPHT-RR-008.0106}
\title{CP asymmetry and branching ratio  of $B\to \pi\pi$}
\author{
E. Kou$^{1}$\footnote{email address: ekou@fyma.ucl.ac.be} and 
T. N. Pham$^{2}$\footnote{email address: Tri-Nang.Pham@cpht.polytechnique.fr}}
\address{
$^1$~Institut de Physique Th\'{e}orique, Universit\'{e} Catholique de Louvain, B1348, Belgium \\
$^2$ Centre de Physique Th\'{e}orique, Centre National de la Recherche Scientifique, UMR 7644 \\ 
Ecole Polytechnique, 91128 Palaiseau, Cedex, France }
\date{\today}
\begin{abstract}
We investigate the branching ratios and  CP asymmetries of the $B\to \pi\pi$
processes measured in B factory experiments. Fits to the experimental data
of this process indicate a large ratio of color-suppressed ($C$) to
color-allowed ($T$) tree contributions. We investigate whether the 
large $C/T$ can be explained within the QCD based model computation with 
i) a large effect from the end-point singularity or with ii) large 
final-state-interaction phase between two different isospin amplitudes. 
We show that the current experimental data do not exclude either 
possibility but we may be able to distinguish these two effects 
in future measurements of direct CP asymmetry of $\pzpz$. 
\end{abstract}
\maketitle
\section{INTRODUCTION}
Recent measurements of the branching ratio and CP asymmetry of the $B\to
\pi\pi$ process provide us with a deep insight into the nature of both
weak and strong interactions. The measurement of the direct  CP
asymmetry in $\pppm$ clearly indicate that there is a substantial
contribution from the $b\to d$ penguin-loop diagram in addition to 
 the dominant $b\to u$ tree-level diagram, which considerably
 complicates the extraction of the  weak phase $\alpha (\phi_2)$ from this 
process. Furthermore, new physics contributions to this penguin diagram 
are not yet excluded. Although  the $B_d-\overline{B}_d$ oscillation
measurement constrains very strictly the new physics contribution to the
$b\to d$ transition coming from the box  diagram,  the one-loop penguin 
diagrams could still get additional contributions in  various  new physics 
models (see \cite{bsbs} for an example).  On the other hand, the biggest 
challenge in the analysis of the $B\to \pi\pi$ processes lies in the 
difficulty of estimating the relative sizes of different topologies, 
which are governed not only by weak interactions but also by  
strong interactions. Therefore, an understanding of the strong
interaction effects in these processes is crucial for  extracting
 the weak phase and ultimately, possible new physics contributions. 

Recently, the combined analysis of the  CP asymmetries of $\pppm$ and  
the branching 
ratios of $\pppm$, $\pzpz$ and $\pppz$ showed  interesting results
for the relative sizes of different types of the tree diagrams. 
At the leading order in QCD, the ratio of the color-suppressed to the 
color-allowed tree diagram, which we call $C/T$, is $1/N_c$, where
$N_c$  is the number of colour, i.e $N_c=3$ in QCD. On the contrary, various
model-independent analysis 
of experimental data  indicates however $C/T$ is close to 
unity \cite{Gronau:2002qj}-\cite{Raz:2005hu}.  
We here would like to investigate whether this large value 
of $C/T$ can be explained by  higher order QCD corrections or  
other hadron dynamics. 

In this article, we investigate two possible enhancement factors of
$C/T$, i) the higher order correction of the QCD based 
model (QCD factorization) \cite{Beneke:2000ry}, \cite{Beneke:2003zv} 
and ii) the effect of FSI phase. For i), we present
an  anatomy of the higher order QCD corrections and discuss in detail, 
the effect of the free parameters using the 
$c$-convention \cite{Gronau:2002qj}. We also show that $C/T$ in QCD
factorization, in the  $c$-convention which we use in this analysis, 
contains contributions from top- and up-penguin as well as annihilation 
diagrams in addition to the pure
color-suppressed tree diagrams. According to QCD factorization, 
these annihilation terms which suffer from the end-point singularity and 
contain free parameters could play an important role in
the enhancement of the $C/T$ ratio. Estimate of annihilation contributions in 
QCD sum-rule can be found in \cite{melic}. 
For ii),  it was found 
in \cite{Cheng:2004ru} that $C/T$ can be {\it effectively} enhanced by 
including non-zero FSI phase. We examine this possibility in 
detail. In this analysis, we use a ``bare'' $C/T$ ratio estimated 
from QCD factorization but 
by suppressing the strong phase from the perturbative computation. 
The other approach including both perturbative and FSI phases can be found in \cite{kim}. 

The remaining of the article is organised as follows. In 
section \ref{sec:fit}, we  fit  the experimental data to a  
model independent parameterization. In section \ref{sec:qcd}, we show the 
prediction of QCD factorization for the parameters defined in 
section \ref{sec:fit}. In section \ref{sec:fsi}, we introduce the
 FSI phase based on the isospin decomposition of the amplitude and show
 how large $(C/T)_{\eff}$ can get to. And finally, we conclude in 
section \ref{sec:concl}. 
\section{MODEL INDEPENDENT FIT OF EXPERIMENTAL DATA }\label{sec:fit}
In this section, we first introduce a model independent parameterization 
for the amplitudes of the $B\to \pi\pi$ processes and summarize the 
fitted values of these parameters to the experimental data. 
\begin{table*}[t!]
\begin{ruledtabular} 
\begin{tabular}{|c||c|c|c|c|c|c|c|}\hline
\ \ \ \ \ \ \ \ \ \ \ \ \ $\gamma$ &&&&&&&\\
$(S_{\pi^+\pi^-}, C_{\pi^+\pi^-})$ \ \ \  &$27^{\circ}$&$37^{\circ}$&$47^{\circ}$&$57^{\circ}$&$67^{\circ}$&$77^{\circ}$&$87^{\circ}$\\
\hline\hline
 & 0.53&0.38&0.36&0.46&0.63&0.81&0.98\\
(-0.62, -0.47)&$-155^{\circ}$&$-131^{\circ}$&$-92.6^{\circ}$&$-61.7^{\circ}$&$-45.0^{\circ}$&$-35.5^{\circ}$&$-29.5^{\circ}$\\ 
& 0.43&0.75&1.10&1.45&1.75&1.96&2.06 \\ \hline 
&0.49&0.29&0.19&0.32&0.52&0.72&0.91\\
(-0.62, -0.27)&$-166^{\circ}$&$-147^{\circ}$&$-92.1^{\circ}$&$-43.4^{\circ}$&$-27.5^{\circ}$&$-20.5^{\circ}$&$-16.6^{\circ}$\\ 
&0.40&0.69&1.03&1.35&1.63&1.82&1.91 \\ \hline
& 0.55&0.38&0.26&0.29&0.44&0.62&0.80 \\ 
(-0.50, -0.37) &$-164^{\circ}$&$-149^{\circ}$&$-115^{\circ}$&$-66.6^{\circ}$&$-41.3^{\circ}$&$-29.8^{\circ}$&$-23.4^{\circ}$ \\
&0.35&0.62&0.92&1.21&1.46&1.63&1.72 \\  \hline 
 &0.61&0.45&0.33&0.31&0.41&0.56&0.72 \\ 
(-0.38, -0.47) &$-164^{\circ}$&$-150^{\circ}$&$-125^{\circ}$&$-86.9^{\circ}$&$-56.7^{\circ}$&$-40.4^{\circ}$&$-31.2^{\circ}$ \\ 
&0.33&0.58&0.85&1.12&1.35&1.51&1.58 \\ \hline 
 &0.59&0.40&0.23&0.17&0.31&0.49&0.67\\ 
(-0.38, -0.27)&$-170^{\circ}$&$-162^{\circ}$&$-140^{\circ}$&$-79.1^{\circ}$&$-37.8^{\circ}$&$-24.1^{\circ}$&$-17.8^{\circ}$ \\ 
&0.31&0.55&0.81&1.07&1.28&1.44&1.51 \\ \hline
\end{tabular}\end{ruledtabular} 
\caption{Determination of $P/T$ (upper value), $\delta_{PT}$ (middle value) and $R$ (bottom value) 
using experimental results for $S_{\pi^+\pi^-}=(-0.50\pm 0.12)$ and $C_{\pi^+\pi^-}=(-0.37\pm 0.10)$ for given values of $\gamma$, $\gamma=(27^{\circ}\sim 87^{\circ})$ . } 
\end{table*}
Let us start by giving the amplitudes of the $B\to \pi\pi$ processes in 
terms of $T$ (color-allowed tree), $C$ (color-suppressed tree), 
$P$ (penguin), which correspond to different topologies, which we 
discuss later on: 
\bea
{\rm Amp} (B^0\to \pi^+\pi^-)&=&T e^{i\delta_t}e^{i\gamma}+P e^{i\delta_P}  \label{eq:1}\\
\sqrt{2}{\rm Amp} (B^0\to\pi^0\pi^0)&=&C e^{i\delta_C}e^{i\gamma}-P e^{i\delta_P} \label{eq:2}\\
\sqrt{2}{\rm Amp} (B^+\to \pi^+\pi^0)&=&(T e^{i\delta_T}+C e^{i\delta_C})e^{i\gamma},  \label{eq:3}
\eea
where $\delta_i$'s is the strong phase and $\gamma$ is the CP violating phase. 
In the following, we analyse 5 observables of the $B\to \pi\pi$ process, 
which are experimentally found to be \cite{HFAG}
 \bea
S_{\pi^+\pi^-}&=&-0.50\pm 0.12 \label{eq:data1}\\
C_{\pi^+\pi^-}&=&-0.37\pm 0.10 \\
{\rm Br}(\pi^+\pi^-)&=&(4.5\pm 0.4)\times 10^{-6} \label{eq:data3}\\ 
{\rm Br}(\pi^0\pi^0)&=&(1.45\pm 0.29)\times 10^{-6} \\ 
{\rm Br}(\pi^+\pi^0)&=&(5.5\pm 0.6)\times 10^{-6} \label{eq:data5} 
 \eea 
where 
${\rm Br}(f_1f_2)$ represents the  CP-averaged branching ratios, ${\rm Br}(f_1 f_2)=(Br(B\to f_1f_2)+Br(\bar{B}\to \bar{f}_1\bar{f}_2))/2$. 
The time-dependent CP asymmetry of $B\to \pi^+\pi^-$ is defined as
\bea
A_{\pi^+\pi^-}(t) &=& 
\frac{\Gamma_{\bar{B} (t)\to \pi\pi}-\Gamma_{B(t)\to\pi\pi}}{\Gamma_{\bar{B}(t)\to \pi\pi}+\Gamma_{B(t)\to\pi\pi}} \\
&=& S_{\pi^+\pi^-}\sin (\Delta M_B t)-C_{\pi^+\pi^-}\cos (\Delta M_B t) \no
\eea
where
\be
S_{\pi^+\pi^-} = \frac{2Im\left(\frac{q}{p}\bar{\rho}_{\pi^+\pi^-}\right)}{1+|\bar{\rho}_{\pi^+\pi^-}|^2}, 
\quad 
C_{\pi^+\pi^-} = \frac{1-|\bar{\rho}_{\pi^+\pi^-}|^2}{1+|\bar{\rho}_{\pi^+\pi^-}|^2}
\ee
with $\bar{\rho}={\rm Amp}(\overline{B}^0\to \pi^+\pi^-)/{\rm Amp}(B^0\to \pi^+\pi^-)$ and $|B_{1,2}\rangle =p|B^0\rangle\pm q|\overline{B}^0\rangle$. In the standard model, we have 
$\frac{q}{p} =\frac{V^*_{tb}V_{td}}{V_{tb}V^*_{td}} =e^{-2i\beta}$ and $\beta (\phi_1)$ is measured in a very high precision from the time-dependent CP asymmetry of $B\to J/\psi K_S$. Using \eq{eq:1}, we obtain 
\be
\bar{\rho}(\pi^+\pi^-)=\frac{T e^{i\delta_T}e^{-i\gamma}+Pe^{i\delta_P}}{T e^{i\delta_T}e^{i\gamma}+Pe^{i\delta_P}} 
\ee
and then, using $\alpha+\beta+\gamma=\pi$, we find (find more detailed derivation, e.g. in \cite{Gronau:2002qj}),  
\bea
&& R\ S_{\pi^+\pi^-}\label{eq:S}\\
&& =\sin 2\alpha +2\sin(\beta-\alpha)\cos\delta_{PT}\left(\frac{P}{T}\right)-\sin 2\beta \left(\frac{P}{T}\right)^2 \no \\ 
 &&R\ C_{\pi^+\pi^-}
 \label{eq:C} \\
&& =2\sin (\alpha+\beta)\sin\delta_{PT} \left(\frac{P}{T}\right)  \no
  \eea 
 where 
 \be
 R=1-2\cos(\alpha+\beta)\cos\delta_{PT}  \left(\frac{P}{T}\right) + \left(\frac{P}{T}\right) ^2.  \label{eq:R}
 \ee
 \begin{figure*}[t]
\begin{center}
\psfrag{ct}[c][c][1]{$C/T$}\psfrag{dct}[c][c][1]{$\delta_{CT}$}
\psfrag{rn}[c][c][.8]{\ $R_{00}$}\psfrag{rc}[c][c][.8]{\ $R_{+-}$}
\psfrag{gamma}[c][c][0.8]{$\gamma =$}
\includegraphics[width=6cm]{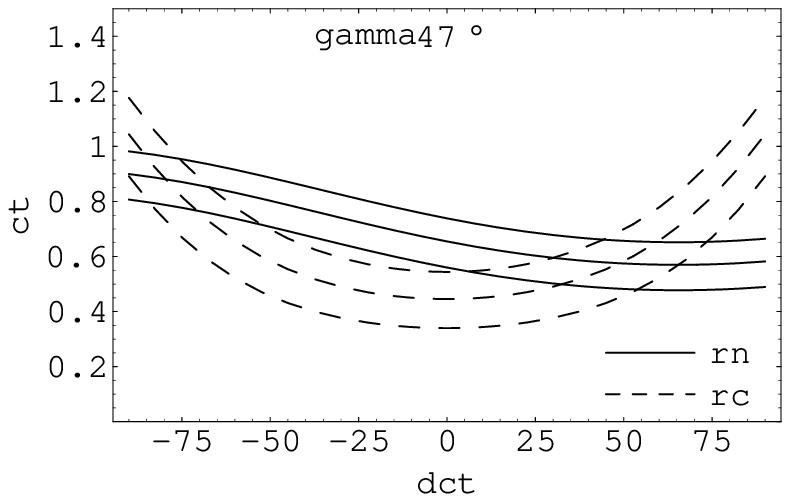}\hspace*{1cm}
\includegraphics[width=6cm]{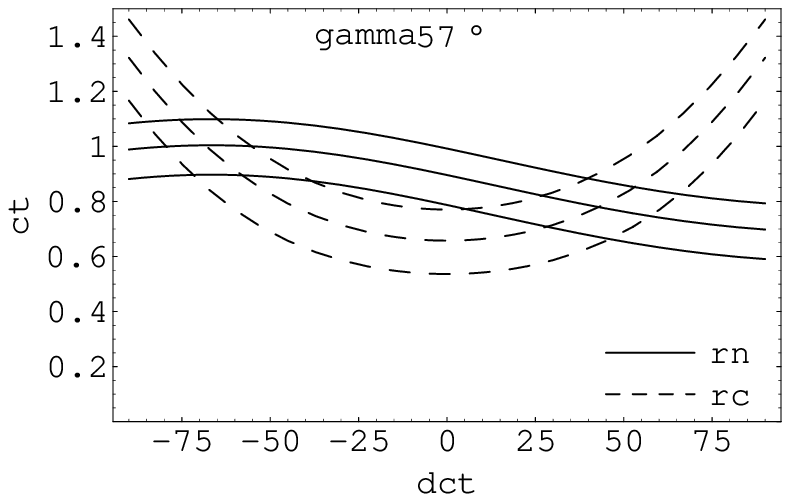} \vspace*{.5cm}\\
\includegraphics[width=6cm]{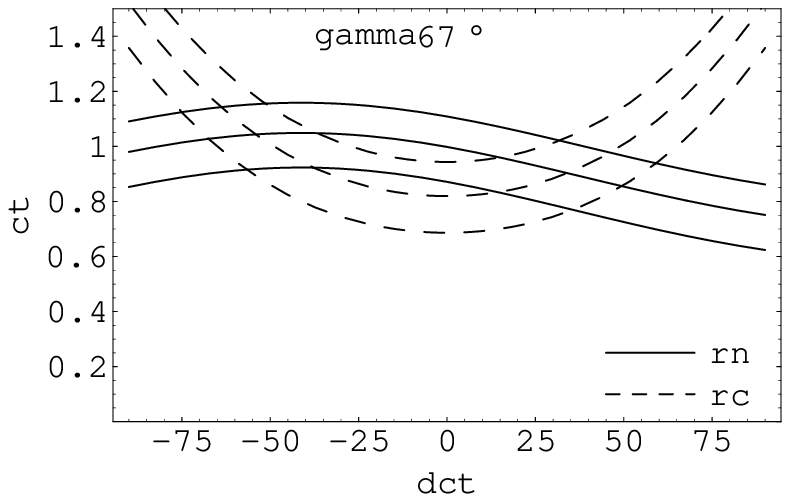}\hspace*{1cm}
\includegraphics[width=6cm]{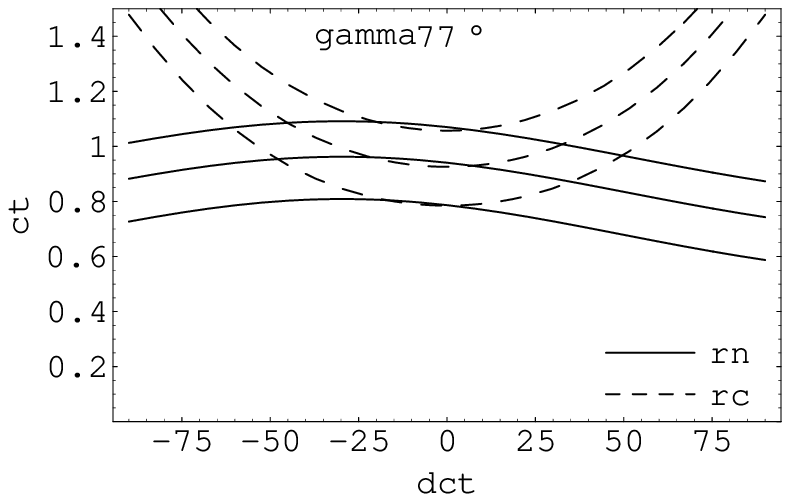}
\caption{Allowed region for $\delta_{CT}$ ($x$-axis) versus $C/T$ ($y$-axis), obtained from experimental bounds for $R_{00}$ and $R_{+-}$. The numerical results for  $P/T$ and $\delta_{PT}$ obtained from the central value of the asymmetry measurements, $(S_{\pi^+\pi^-}, C_{\pi^+\pi^-})=(-0.50, -0.37)$ (see Table 1) are used. 
The three solid lines represent $R_{00}=0.64-0.14,\ 0.64, \ 0.64+0.14, $
and the three dashed lines represent  $R_{+-}=2.27-0.32,\ 2.27, \ 2.27-0.32$.  
The overlap of solid and dashed bounds are the allowed region for $C/T$ and $\delta_{CT}$. 
The weak phase $\gamma$ is fixed as left-top ($\gamma=47^{\circ}$), 
 right-top ($\gamma=57^{\circ}$),  left-bottom ($\gamma=67^{\circ}$),  right-bottom ($\gamma=77^{\circ}$). }
\label{fig:1}
\end{center}
\end{figure*}
As for the branching ratios, we follow \cite{Buras:2003dj} and use the ratios of the averaged branching ratios, which are derived from \eq{eq:1}, \eqeq{eq:2}{eq:3} as  
 \bea
R_{00}&=&\frac{2{\rm Br}(\pi^0\pi^0)}{{\rm Br}(\pi^+\pi^-)}\label{eq:rn}\\
 &=&\frac{1}{R}\left[ \left(\frac{C}{T}\right)^2 + \left(\frac{P}{T}\right)^2 \right. \no \\
 && \left. -2\cos(\delta_{PT}-\delta_{CT})\cos\gamma \left(\frac{C}{T}\right)  \left(\frac{P}{T}\right)\right]
 \no \\
R_{+-}&=&\frac{2{\rm Br}(\pi^+\pi^0)\tau_{B^0}}{{\rm Br}(\pi^+\pi^-)\tau_{B^+}} \label{eq:rc} \\
 &=&\frac{1}{R}\left[ 1 +2\cos\delta_{CT}\left(\frac{C}{T}\right)+\left(\frac{C}{T}\right)^2  \right]
 \no 
 \eea
 where $\delta_{ab}\equiv \delta_a-\delta_b$. 

Before discussing our result, we would like to make a comment on
the direct CP asymmetry of the $\pi^0\pi^0$ channel, $C_{00}$. In the
same parameterization, one can write: 
\be
C_{00}=\frac{2\sin\gamma\sin (\delta_{CT}-\delta_{PT})\big(\frac{C}{T}\big)\big(\frac{P}{T}\big)}{\big(\frac{C}{T}\big)^2+\big(\frac{P}{T}\big)^2-2\cos\gamma\cos (\delta_{CT}-\delta_{PT})\big(\frac{C}{T}\big)}.  
\ee
The experimental bound is given as \cite{HFAG}: 
\be
C_{00}=0.28^{+0.40}_{-0.39}. \label{eq:c00}
\ee
Since the experimental data is not very precise yet, we will not include
this data in our analysis but will discuss its relevance to the
strong phase $\delta_{CT}$ in subsequent sections.

 Now using these formulae, we shall fit the parameters to the 
experimental data. 
Table 1 shows determinations  of $P/T$ (upper values) , $\delta_{PT}$ 
(middle values) and $R$ (bottom values) by using experimental values 
of $S_{\pi^+\pi^-}$ and $C_{\pi^+\pi^-}$ for given values of $\gamma$, by using $\beta=23.7^{\circ}$. 
We can find that the $R$ value becomes larger than unity in the most of 
the parameter space for $\gamma > 57^{\circ}$. We also find that $R$ is 
particularly larger when $S_{\pi^+\pi^-}$ is larger and negative. 

Next inputting the values of  $P/T$ and $\delta_{PT}$ obtained from 
the above analysis into the R.H.S. of Eq. (\ref{eq:rn}) and Eq. (\ref{eq:rc}) 
and the experimental values of $R_{00}$ and $R_{+-}$ into the L.H.S., 
we compute  
$C/T$ and $\delta_{CT}$. We first use only the central value of 
 $(S_{\pi^+\pi^-}, C_{\pi^+\pi^-})=(-0.50, -0.37)$ but include 1$\sigma$ experimental error for 
 $R_{00}$ and $R_{+-}$. 
Obtained results for $\gamma=47^{\circ}$ (left-top),\
$57^{\circ}$ (right-top),\  $67^{\circ}$ (left-bottom),\  $77^{\circ}$ 
(right-bottom) are shown in Fig. 1. 
The overlap of the solid ($R_{00}$) and the dashed ($R_{+-}$)  bounds shift towards the larger $C/T$ region as $\gamma$ becomes larger, or equivalently $R$ becomes larger. 
Therefore, the large value  of $R$, which is originated from the large negative $S_{\pi^+\pi^-}$, causes the large value of $C/T$. 
Furthermore, we find that $R_{+-}$ allows relatively small value of $C/T$ 
while $R_{00}$ leads to a more strict constraint, $C/T \gsim 0.5$. We find that the overlap region is distributed in a large range of $\delta_{CT}$.  
Let us now discuss the errors coming from $(S_{\pi^+\pi^-}, C_{\pi^+\pi^-})$, since 
Fig. 1 is obtained by using only their central values.  
First, both $R_{00}$ and $R_{+-}$ depend on $(S_{\pi^+\pi^-}, C_{\pi^+\pi^-})$ through $R$, 
as $1/R$ as mentioned above.   
While $R_{+-}$ does not have further $P/T$ and $\delta_{pt}$ dependence,  
$R_{00}$ has more complex dependence on them.  However, as long as the overlap region is concerned, we find that the derived error is up to $\pm$ a few \% in $C/T$ and  $\pm 20^{\circ}$ in $\delta_{CT}$. 

From the above analysis, we can not  obtain a strong constraint on the
weak phase $\gamma$. While measurements for e.g. the direct CP asymmetry
of the $\pi^0\pi^0$ channel would allow a determination of
$\gamma$ in the future, currently, we need some inputs from 
theoretical models. 
Especially, the value of $C/T$ found from the fits in this section seems
to be rather large comparing to the  leading order prediction. Therefore, we will try to extract bounds for $C/T$ and $\delta_{CT}$ using the theoretical models in the following section, which in turn may give us a constraint on $\gamma$. 
\section{QCD MODEL CALCULATION OF PARAMETERS $C,T$ AND $P$}\label{sec:qcd}
\begin{table*}[t]
\begin{ruledtabular}
\begin{tabular}{|c||c|c|c|c|}\hline
 &factorisable & vertex corr. & hart-scat. corr. & penguin corr. \\ \hline\hline
$a_1$  & 1.02 (1.04)&$0.032 e^{i 27^{\circ}} (0.044 e^{i 42^{\circ}})$&$-0.032-0.014 \rho_H e^{i \phi_H} (-0.061-0.025 \rho_H e^{i\phi_H})$&0 (0)\\ \hline 
$a_2$&$0.17 (0.085)$&
$-0.18 e^{i 27^{\circ}} (-0.19 e^{i 42^{\circ}})$&$0.18 +0.081\rho_H e^{i \phi_H} (0.24+0.095 \rho_He^{i\phi_H})$&0 (0) \\ \hline
$a_4^u$&$-0.031 (-0.046)$&
$-0.0023 e^{i 27^{\circ}} (-0.0034 e^{i 42^{\circ}})$&$0.0023+0.0010\rho_H e^{i \phi_H} (0.0047+0.0019 \rho_He^{i\phi_H})$&$0.014 e^{-i 73^{\circ}} (0.022e^{-i50^{\circ}})$ \\ \hline
$a_4^c$&$-0.031 (-0.046)$&
$-0.0023 e^{i 27^{\circ}} (-0.0034 e^{i 42^{\circ}})$&$0.0023+0.0010\rho_H e^{i \phi_H} (0.0047+0.0019 \rho_He^{i\phi_H})$&$-0.0047 e^{i 76^{\circ}} (0.0084e^{-i27^{\circ}})$ \\ \hline
$a_6^u$&$-0.039 (-0.060)$&
$-0.00047 (-0.00083) $&$0 (0)$&$-0.014 e^{i 79^{\circ}} (0.017e^{-i 73^{\circ}})$ \\ \hline
$a_6^c$&$-0.039 (-0.060)$&
$-0.00047 (-0.00083) $&$0 (0)$&$-0.0073 e^{i 38^{\circ}} (0.0038 e^{-i 78^{\circ}})$ \\ \hline
\end{tabular}\end{ruledtabular}
\caption{Anatomy of the higher order correction in the QCD factorization. For the input paramters, we use the central values given in \cite{Beneke:2003zv}. The numbers in the parenthesis are obtained by changing renormalisation scale to $\mu=2.1$ GeV from the default value. } 
\end{table*}

In this section, we  investigate whether those fitted values on $C/T$ 
and $\delta_{CT}$ can be reproduced by the QCD factorization. 
Let us first give the relation between the parameterisation of the amplitudes in Eqs. (\ref{eq:1}) to (\ref{eq:3}) in the previous section  and the one in QCD factorization: 
\bea
Te^{i \delta_T}e^{\gamma}&\propto&\lambda_u^*(a_1+b_1+\hat{a}^u_4) , \label{eq:Tqcd}\\ 
Ce^{i \delta_C}e^{\gamma}&\propto&\lambda_u^*(a_2-b_1-\hat{a}^u_4) , \label{eq:Cqcd}\\ 
Pe^{i \delta_P}&\propto&\lambda_c^*\hat{a}^c_4 \label{eq:Pqcd}
\eea
where 
\be
\hat{a}_4^p=a_4^p+r_\chi a_6^p+2b_4+b_3
\ee
and $r_\chi=2m_{\pi}^2/(2m_b m_q)\simeq 1.24$ with $m_q\equiv (m_u+m_d)/2$.
Here we employ the so-called $c$-convention, which eliminates
 $\lambda_t$ by using an unitarity relation. Therefore, the amplitudes 
are proportional  only to two   CKM factors $\lambda_u $ and $\lambda_c $
\bea
\lambda_u=V_{ub}V_{ud}^*\simeq A\lambda^3(\rho -i\eta ) \label{eq:lambdau}\\
 \lambda_c=V_{cb}V_{cd}^* =\simeq -A\lambda^3. \label{eq:lambdac}
\eea
Note that $\arg (\rho-i\eta )=e^{i\gamma}$ and 
$|\rho-i\eta |=|\lambda_u^*/\lambda_c^*|$. 
It is important to notice that apart from the ``pure'' color-allowed tree
contribution $a_1$ and the ``pure" color-suppressed tree contribution 
$a_2$, $Te^{i \delta_T}$ and $Ce^{i \delta_C}$ contain the same two
terms with opposite sign, which are penguin- and tree-annihilations 
contributions ($b_i$ terms) and top- and up-penguin contributions 
 ($a_4^u$). As has already been investigated in \cite{Beneke:2003zv}, it is 
quite possible that these contributions could effectively enhance the 
ratio $C/T$ by  contributing constructively and destructively  to 
$C$ and $T$, respectively.  In this respect, the sign of these extra 
contributions must be carefully investigated.  
\begin{figure*}[t]
\begin{center}
\psfrag{ct}[c][c][1]{$C/T$}\psfrag{dct}[c][c][1]{$\delta_{CT}$}
\includegraphics[width=5.5cm]{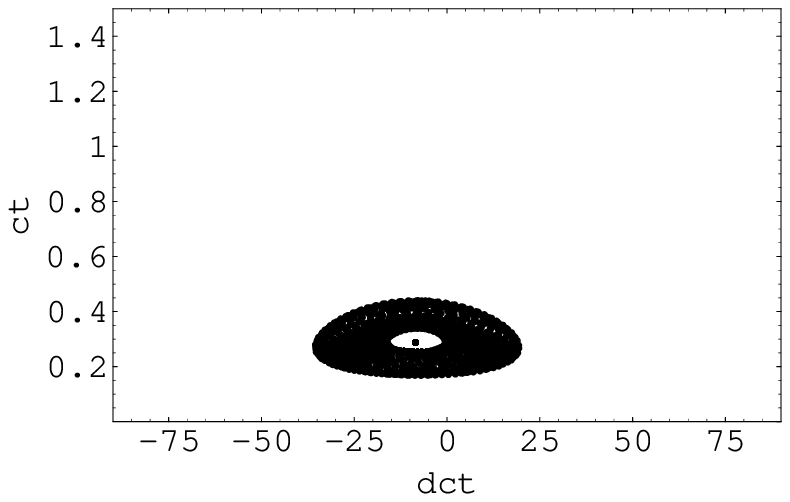}\hspace*{0.3cm}
\includegraphics[width=5.5cm]{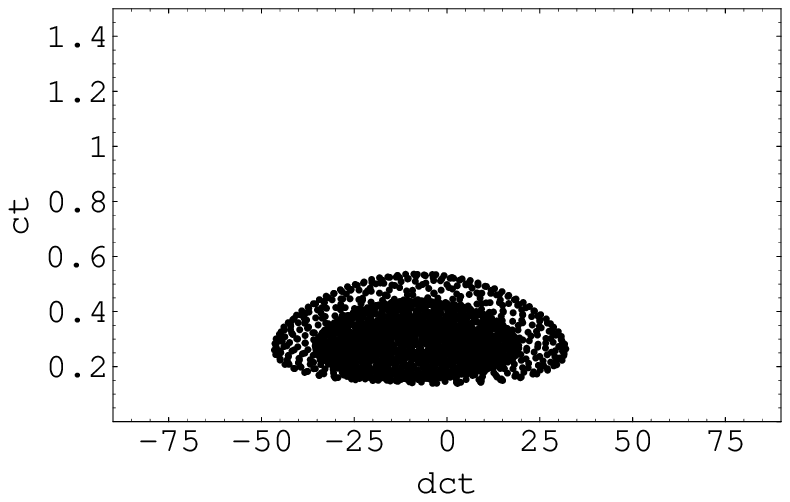}\hspace*{0.3cm}
\includegraphics[width=5.5cm]{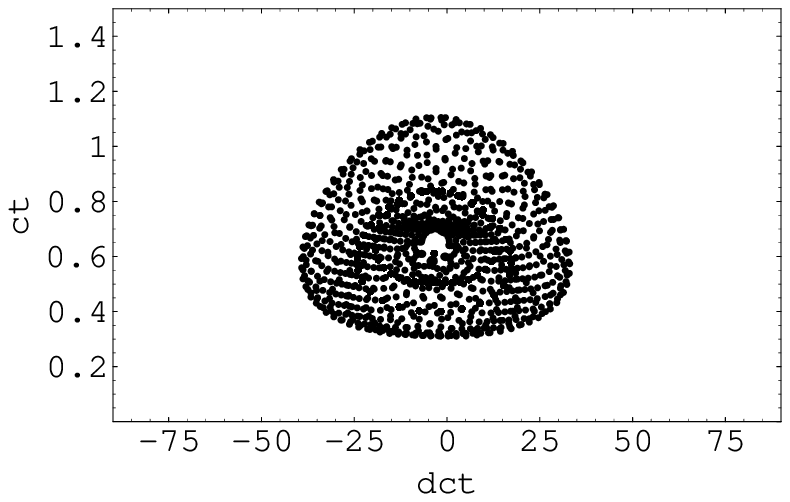}
\caption{Scattered plot of the QCD factorization estimate for $\delta_{CT}$ ($x$-axis) versus $C/T$ ($y$-axis)  including the end-point singularity effects. In the plot, we fix the $\rho$ parameters as $(\rho_H,  \rho_A)=(1, 1)$ (left) and $=(1, 2)$ (middle) and vary the phases in the range of  $-\pi < \phi_{A, H}<\pi$ (interval of 0.2 radian). The rest of the parameters are fixed (see text for details). The last figure (right) is obtained in the same manner with $(\rho_H,  \rho_A)=(1, 1)$ but with different parameter set, the so-called scenario 2 of QCD factorization (see text for details). }\label{fig:3}
\end{center}
\end{figure*}

In order to understand the  size of the higher order corrections  
estimated by the QCD factorization, we first give  an expression 
decomposing $a_i^{p}$ and $b_i$ into factorisable terms and 
their correction terms: 
\bea
 a_i^p&=&\left(C_i+\frac{C_{i\pm 1}}{N_c}\right) +\frac{C_{i\pm 1}}{N_c} \frac{C_F \alpha_s}{i\pi}
\left[V_i +\frac{4\pi^2}{N_c}H_i\right] +P_i^p \no \\ && \\
b_1&=&\frac{C_F}{N_c^2}C_1 A_1^i \\
b_3&=&\frac{C_F}{N_c^2}[C_3 A_1^i+C_5(A_3^i+A_3^f)+N_cC_6A_3^f] \\
b_4&=&\frac{C_F}{N_c^2}[C_4 A_1^i+C_6A_2^i]
\eea
where $p=u, c$. The sign $\pm$ in $a_i$ must be taken as $+$ for $i=\mbox{odd}$ and $-$ for $i=\mbox{even}$. 
The first terms of $a_i^{p}$ are called  factorisable term. The term
proportional to $V_i, H_i, P_i^{p}, A_i^{i, f}$ are the vertex
correction, hard-scattering correction, penguin correction and
annihilation correction, respectively. At the leading order,   all the
Wilson coefficients vanish  except  $C_1$ with $C_1=1$, which leads to
\be
C/T=1/3, \ \ \ \ \ \ \ P/T=0,  \ \ \ \ \ \ \ \ \mbox{At LO}.  
\ee
The numerical results including all the above higher order corrections are shown in Table 2. 
For the input parameters, we use the central values in the Table 1 of \cite{Beneke:2003zv}, among which we list some important ones here; 
\bea
&\mu=4.2  \mbox{GeV}, \ \ \ m_q(2\mbox{GeV})=0.0037  \mbox{GeV}, \ \ \ \lambda_B=0.35 \mbox{GeV}, & \no \\
&|\lambda_u/\lambda_c|=|\rho +i\eta |= 0.09, \ \ \ \alpha_2^{\pi}=0.1 & 
\eea
where $\lambda_B$ and $\alpha^{\pi}_2$ are the parameters for the distribution amplitude of $B$ meson and $\pi$, respectively (for theoretical estimates of these parameters, see e.g. \cite{Ball:2003fq}-\cite{Lee:2005gz} and \cite{Ball:2005tb}). The value of $m_q$  must be running to the appropriate scales in the computation. The numbers in the parenthesis in Table 2 are the results with a smaller  renormalisation scale, $\mu=2.1$ GeV (the other parameters are the same as before). 
We can see that  the Wilson coefficients $C_{2- 6}$ are  $O(\alpha_s)$ suppressed comparing to $C_1$ and  $a_1$ is completely dominated by the factorisable term. On 
the other hand, the factorisable term of $a_2$ is rather small 
since $C_2$ is  $O(\alpha_s)-$ suppressed and there is a color 
factor $1/N_c$ in  $C_1$ term and furthermore, these two have opposite
signs. As a result, the higher order corrections, $V_2$ and $H_2$ terms, which 
are proportional to the leading order Wilson coefficient $C_1$, lead to 
large  contributions in $a_2$. It is also important to notice that these 
correction terms can induce a large strong phase in $a_2$ which has
a comparable real and imaginary part in contrast to $a_{1}$ which is
almost real. In fact, in 
the soft-colliner effective theory (SCET) \cite{Bauer:2004tj}
-\cite{Bauer:2005kd}, this correction to $a_2$ which is proportional to
a large coefficient $C_1$ contains some free parameters.  
So it could  be much more enhanced in SCET; as much as solving the 
problem of large $C/T$.   A more recent analysis in SCET can also be found in \cite{zupan}. 
The smaller $\mu$ value reduces the factorisable term of $a_2$ and thus, the $C/T$ value. 
We should also mention that the penguin terms $a_{4 (6)}^{u}$ 
and $a_{4 (6)}^{c}$ are quite similar apart from penguin correction
terms. The difference in the penguin corrections is due to 
charm- and up-penguin difference.  
Using the results with the default renormalisation scale, $\mu=4.2$ GeV, we find 
\bea
\kern -0.5cm a_1&=&1.02 e^{i 0.8^{\circ}}-0.014 \rho_H e^{i \phi_H} \label{eq:a1} \\ 
\kern -0.5cm a_2&=&0.21e^{-i 23^{\circ}}+0.081  \rho_H e^{i \phi_H} \label{eq:a2} \\
\kern -0.5cm a_4^u+r_\chi a_6^u&=&
- 0.097e^{i 21^{\circ}}+0.0010  \rho_H e^{i \phi_H}\label{eq:a4u} \\
a_4^c+r_\chi a_6^c&=&
- 0.10e^{i 7^{\circ}}+0.0010  \rho_H e^{i \phi_H}  \label{eq:a4c}
\eea
and 
\bea
\kern -0.5cm b_1&=&0.027+0.063 \rho_A e^{i \phi_A} +0.0085 (\rho_A e^{i \phi_A})^2  \label{eq:b1} \\
\kern -0.5cm b_3&=&-0.0067-0.021 \rho_A e^{i \phi_A} -0.015 (\rho_A e^{i \phi_A})^2  \label{eq:b2} \\
\kern -0.5cm b_4&=&-0.0019-0.0046 \rho_A e^{i \phi_A} -0.00061 (\rho_A e^{i \phi_A})^2. \label{eq:b3} 
\eea
The parameters $\rho_{A, H}$ and $\phi_{A, H}$ which originate from the 
end-point singularity would vary, say, in the ranges of  
$|\rho_{A,  H}|<1\sim 2$ 
and $-\pi < \phi_{A, H}<\pi$. We  find that 
$\rho_H$ and $\phi_H$ have significant contributions to $a_2$, 
i.e. $C$ and $\rho_A$ and $\phi_A$ to $b_i$, i.e. all of $T, C, P$. 
According to \eqeq{eq:Tqcd}{eq:Cqcd}, in the $c-$ convention,
$C/T$ is not simply $a_2/a_1$ 
but includes extra contributions from $\hat{a}^u_4$ and $b_1$, which, 
we find,  are as large as $a_2$ and strongly depend on $\rho_A$ 
and $\phi_A$. We perform complete analysis of $C/T$ covering all 
the parameter space of $\rho$'s and $\phi$'s next. Here, however,  it 
is very important to notice that at the limit of $\rho_{H, A}=0$, numerical values of 
$a_{1, 2}$ and $a_{4, 6}^u$ have the opposite sign, which enhances $
C$ and suppress $T$ (see \eqeq{eq:Tqcd}{eq:Cqcd}), i.e. the inclusion 
of  $a_{4, 6}^u$ terms increases  $C/T$. As a result, we obtain: 
\be
\frac{C_0}{T_0}e^{i\delta_{CT_0}}= 0.29 e^{-i8.5{\circ}} \label{eq:ctrho0}
\ee
where the index 0 indicates $\rho_{H, A}=0$. 
 We emphasize once more 
that the signs of $a_{4, 6}^u$ and $a_{4, 6}^c$ must be the same 
unless there is large enhancement factors for $c-$ and/or 
$u-$penguins. And most importantly, the sign of $a_{4,6}^c$ can 
be fixed from determinations of $P$ and $\delta_{PT}$ up to well-known $\lambda_c$ 
factor (see, \eq{eq:Pqcd}). 
 
Next we consider the effect of the end-point singularity, $\rho_{H, A}$ and $\phi_{H, A}$, 
which often cause large theoretical uncertainties in the prediction of 
QCD factorization. 
The behaviour of  $C/T$ when varying freely these four parameters is 
rather complicated. 
In Fig. 2, we show scattered plots of $\delta_{CT}$ ($x$-axis) versus $C/T$ ($y$-axis) varying the parameters in the range of  $-\pi < \phi_{A, H}<\pi$ (interval of 0.2 radian) and fixing $\rho_H=1$ (left; $\rho_A=1$, middle; $\rho_A=2$). 
We can see that quite a large range of $C/T$ and $\delta_{CT}$ are allowed from QCD factorization, $C/T$  up to 0.45 (0.55) for $\rho_A=1 (2)$.
In particular, the value of $C/T$ becomes large at small negative values of $\delta_{CT}$. 
For the case of  $\rho_A=1$ and $\rho_A=2$, we obtain a constraint respectively,   $\gamma \le 44^{\circ} (52^{\circ})$  and $\gamma \le 46^{\circ}$ ($56^{\circ}$) 
 allowing $1\sigma (2\sigma)$ error in the experimental values, $S_{\pi^+\pi^-}, C_{\pi^+\pi^-}, R_{00}, R_{+-}$.

\begin{figure*}[t]
\begin{center}
\psfrag{d0}[c][c][0.7]{$\delta_{20}=0^{\circ}$}
\psfrag{d20}[c][c][0.7]{$\delta_{20}=20^{\circ}$}
\psfrag{d40}[c][c][0.7]{$\delta_{20}=40^{\circ}$}
\psfrag{d60}[c][c][0.7]{$\delta_{20}=60^{\circ}$}
\psfrag{d80}[c][c][0.7]{$\delta_{20}=80^{\circ}$}
\psfrag{dm20}[c][c][0.7]{$\delta_{20}=-20^{\circ}$}
\psfrag{dm40}[c][c][0.7]{$\delta_{20}=-40^{\circ}$}
\psfrag{dm60}[c][c][0.7]{$\delta_{20}=-60^{\circ}$}
\psfrag{dm80}[c][c][0.7]{$\delta_{20}=-80^{\circ}$}\psfrag{ct}[c][c][1]{$C_{\eff}/T_{\eff}$}\psfrag{dct}[c][c][1]{$\delta_{CT_{\eff}}$}
\psfrag{rn}[c][c][.9]{$R_{00}$}\psfrag{rc}[c][c][.9]{$R_{+-}$}
\psfrag{gamma}[c][c][0.8]{$\gamma =$}
\includegraphics[width=5.5cm]{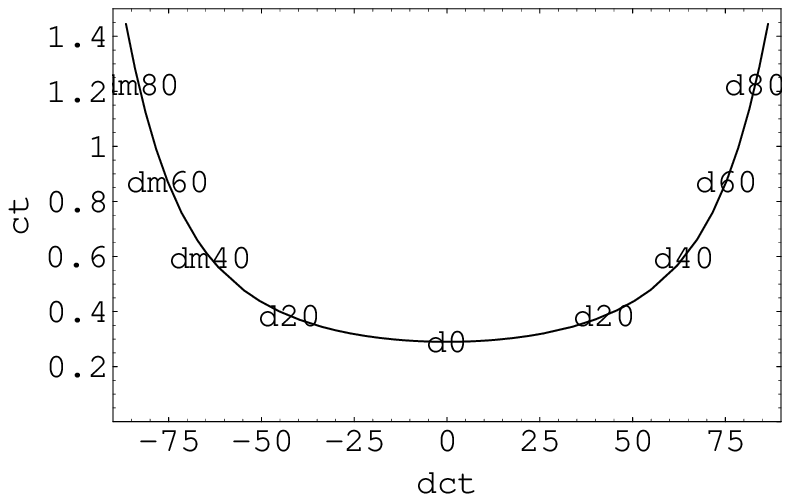}\hspace*{0.3cm}
\includegraphics[width=5.5cm]{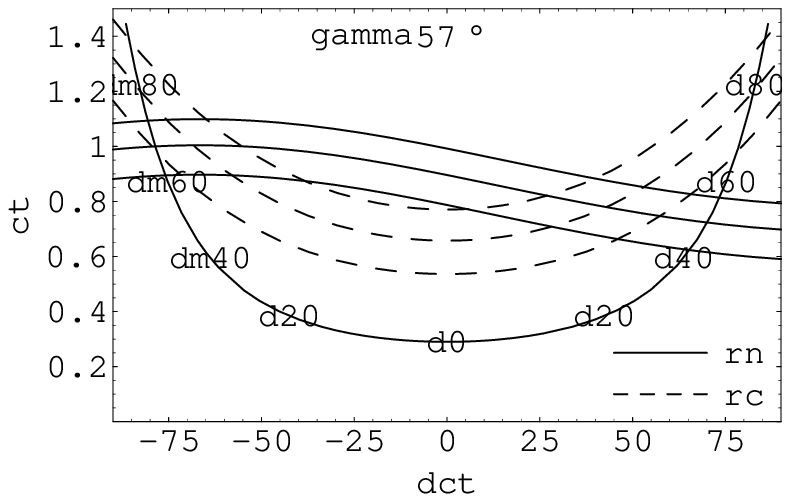}\hspace*{0.3cm}
\includegraphics[width=5.5cm]{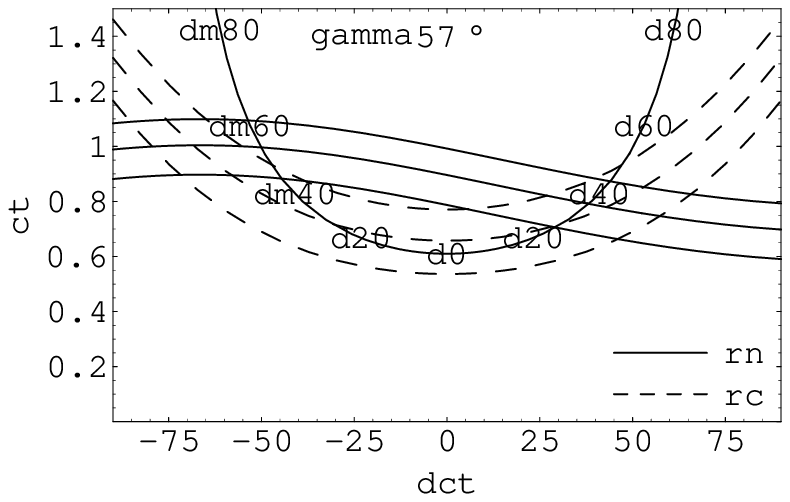}
\caption{The left figure is the plot of Eq. (\ref{eq:ct1}), the result including the FSI phase,  
in the plane of $\delta_{CT_{\eff}}$ ($x$-axis)
 versus  $C_{\eff}/T_{\eff}$ ($y$-axis) by varying $\delta_{20}$. 
 The number on the line indicates the value of $\delta_{20}$ at each point. 
The bare value $C_0/T_0=0.29$  obtained from the default parameter sets of the QCD factorization with $\rho_{H, A}=0$ is used.  In the middle figure, we put together the left figure and the experimental bounds from $R_{+-}$ and $R_{00}$ for the case of $\gamma=57^{\circ}$ (Fig.1 upper-right). The right figure is obtained in the same way as the middle one but using $C_0/T_0\simeq 0.61$, the result with  the parameter set called scenario 2 in QCD factorization. }\label{fig:3}
\end{center}
\end{figure*}

In the original paper of QCD factorization \cite{Beneke:2003zv}, the problem of the small $a_2$ value has already been  recognized and a possible solution was proposed, choosing the largest value of the Gegenbauer moment of $\pi$ distribution amplitude, $\alpha^{\pi}_2\simeq 0.4$ and the smallest value of the first negative moment of the $B$ meson distribution function, $\lambda_B=0.2$ GeV (scenario 2). More recently, this approximation has been reanalysed by using QCD factorization with the 1-loop
(NNLO) corrections to hard spectator-scattering diagram  \cite{beneke2}. 
In this way, the hard-scattering correction is enhanced by a factor of two, which leads to 
\be
a_2\simeq 0.48 e^{-i10^{\circ}}+0.18 \rho_H e^{i \phi_H} \label{eq:s2}
\ee
We found that the effect to  $a_1$ is small since $a_1$ is dominated by the leading order contribution which does not depend on those parameters.  As a result, we obtain $C_0/T_0 e^{i\delta_{CT_0}}\simeq 0.61 e^{-i3^{\circ}}$ for $\rho_{H, A}=0$. 
Note that these lower value of $\lambda_B$ and higher value of $\alpha^{\pi}_2$  must be carefully tested using  other 
charmless $B$ decays which often involve these two 
parameters. Fig. 2 (right most) shows the scattered plot produced as the other Fig 2 but  with $\rho_A=1$ and with parameter set of the scenario 2. We find 
$C/T\simeq 1.1$ can be achieved in this scenario if $\delta_{CT}$ is very small.  
Thus, QCD factorisation can solve the large $C/T$ puzzle. Nevertheless, whether QCD factorisation can reproduce all the data in   Eqs. (\ref{eq:data1}) to (\ref{eq:data5}) simultaneously depends not only on a large $C/T$ but also its prediction on $P/T$ and $\delta_{PT}$, which must be carefully analysed by  comparing e.g. to  penguin dominant modes so as to make sure that  our parameter sets are sensible. 

\section{DOES FSI PHASE MAKE $C/T$ LARGE?}\label{sec:fsi}
In this section, we introduce the FSI phase into our QCD factorization analysis. 
In QCD factorization, this effect is ignored by arguing that the sum of the phases from all possible intermediate states 
cancel each other statistically. This argument has been challenged 
in, e.g.  \cite{Suzuki:1999uc} where it is argued that
this mechanism may work only in the inclusive processes and it is found that the strong phase
in $B\to \pi\pi$ decays can be relatively large. Furthermore, it has been
shown in  \cite{Cheng:2004ru} that the FSI phase can effectively enhance
$C/T$, which is favoured by our analysis in section \ref{sec:fit}. This
is because of isospin invariance; the $B\to \pi^{0}\pi^{0}$ decay can
 be induced by the charge exchange scattering process $\pi^{+}\pi^{-}\to
\pi^{0}\pi^{0}$ which  effectively generates the $C$ amplitude 
from $T$. 
Thus, we consider the case in which the QCD factorization amplitudes contain an additional large FSI phase  between two isospin $I=0,2$ $B\to \pi\pi$ amplitudes, $\delta_{0,2}$, which can generate extra contributions to  $C$ of  the QCD factorization computation. 
We examine whether the FSI effect  can enhance sufficiently the value of $C/T$ of the QCD factorization 
without adjusting the incalculable parameters $\rho_{H,A}$ and $\phi_{H,A}$ coming from the end-point singularities of the annihilation and hard scattering diagrams as shown in section \ref{sec:qcd}. 
For this purpose, we start from the QCD factorization amplitudes with  $\rho_{A,H}=0$ but with the FSI phase $\delta_{2,0}$ and evaluate  $C/T$ and furthermore constrain the values of $\delta_{20}$ and $\delta_{CT}$. Note that we neglect inelastic FSI here.
 While comprehensive computations of
the FSI phase can be found in \cite{Cheng:2004ru} and
\cite{Cheng:2005bg}  followed by \cite{Fajfer:2005mf} (and also in earlier ones \cite{JM} and \cite{isola})   where a large strong phase difference is found, we here examine
these effects in a more phenomenological manner. 
In \cite{WW}, a
similar analysis with strong phases in the isospin amplitudes 
is performed and a large
$\delta_{20}$ is found by a fit to the the central values of the
experimental data. However, as we have seen in section \ref{sec:fit},
the experimental errors are still large to constrain the phase
$\delta_{CT}$ and consequently, the FSI phase, without a theoretical input.

Now, the effective parameters $T_{\eff}, C_{\eff}$, etc..  are related to the parameters in the previous section as 
\bea
\kern -0.5cm T_{\eff} e^{i\delta_{T_{\eff}}} &=& [(2T_0-C_0)e^{i\delta_0}+ (T_0+C_0)e^{i\delta_2}]/3 
\label{eq:27} \\
\kern -0.5cm C_{\eff} e^{i\delta_{C_{\eff}}} &=& [-(2T_0-C_0)e^{i\delta_0}+ 2(T_0+C_0)e^{i\delta_2}]/3 \label{eq:28} \\
\kern -0.5cm P_{\eff} e^{i\delta_{P_{\eff}}}&=&P_{0}e^{i(\delta_D+\delta_{0P})}. \label{eq:29}
\eea
For the parameters $C_0, T_0, P_0$ on the R.H.S., we use the  
QCD factorization prediction with $\rho_{H, A}=0$ following our strategy mentioned above. 
Note that $P_{\eff}$ has not only the $I=0$ phase, $\delta_{0P}$,  
but also an extra phase, $\delta_D$ which may come 
from inelastic re-scattering, such as $DD\to\pi\pi$. 
As a result, the effective color-suppressed to color-allowed ratio 
is obtained as: 
\be
\left(\frac{C_{\eff}}{T_{\eff}}\right)e^{i\delta_{CT_{\eff}}}=\frac{(-2+2e^{i\delta_{20}})+(1+2e^{i\delta_{20}})C_0/T_0}{(2+e^{i\delta_{20}})+(-1+e^{i\delta_{20}})C_0/T_0}\ \ \ \label{eq:ct1}
\ee
The behaviour of this function with the value for $C_0/T_0$ in \eq{eq:ctrho0}  is shown in Fig. 3 (left). We use the 
same  $\delta_{CT_{\eff}}$ (x-axis) versus $(C/T)_{\eff}$ (y-axis)   space shown in Fig. 1. 
The numbers on the line indicates the value 
of $\delta_{20}$ at each point.  We can see that $C/T$ indeed  becomes larger as the FSI phase $\delta_{20}$ increases. We find e.g. that the bare ratio $C_0/T_0=0.29$ can be enhanced to  
$(C_{\eff}/T_{\eff})\simeq 0.4$  for 
$\delta_{20}\simeq \pm 21^{\circ}$, where 
$\delta_{CT_{\eff}}\simeq \pm 44^{\circ}$. 
In Fig. 3 (middle), we   overlap Fig. 3 (left) and the experimental bounds for 
$R_{+-}$ and $R_{00}$ (Fig. 1 for $\gamma=57^{\circ}$). We find that the 
allowed region from $R_{+-}$ and $R_{00}$ overlap at $\delta_{20}\simeq -65^{\circ}$, where $(C/T)_{\eff}\simeq 0.96$. Fig. 3 (right) is obtained
in the same way as the middle figure but with using $C_0/T_0$ value 
from the scenario 2 in section \ref{sec:qcd}. 
We can find that the central value of $(R_{+-}, R_{00})$ are reproduced by 
$\delta_{20}\simeq 40^{\circ}$ where $C_{\eff}/T_{\eff}\simeq 0.8$ and $\delta_{CT_{\eff}}\simeq 40^{\circ}$. 

In order to obtain a constraint on $\gamma$, we further need to know the maximum size of the FSI phase. For example, assuming $\delta_{20}\lsim 30^{\circ}$, we find $\gamma \lsim 48^{\circ} (55^{\circ})$ using the default  values for the input parameters of QCD factorization, i.e. using  \eq{eq:ctrho0} and including $1\sigma (2\sigma)$ of the experimental errors in $S_{\pi^+\pi^-}, C_{\pi^+\pi^-}, R_{00}, R_{+-}$.  However, as we have seen in Fig. 3,  this result depends strongly on the inputs of QCD factorization. For example, with the scenario 2 of section \ref{sec:qcd}, we find that $\delta_{20}\lsim 30^{\circ}$ leads to  $\gamma = (59\pm3)^{\circ} (>56^{\circ})$. It is also important to mention that there may be a FSI contribution not only to the phase but also to $C/T$ itself, as discussed in \cite{Fajfer:2005mf}. Therefore, the bound obtained here may receive a considerable corrections from both uncertainties of QCD factorization and of FSI. Further improvements in estimating those parameters are necessary for obtaining the bound for $\gamma$ from this strategy. 

\section{conclusions}\label{sec:concl}
We analysed the latest measurements of branching ratios and CP asymmetry
in the $B\to \pi\pi$ processes and compared it  to the theoretical model
predictions. Using a  model independent parameterisation of the 
$B\to \pi\pi$ process, we first constrained the penguin-tree ratio
 parameters, $P/T$ and $\delta_{PT}$ by using the asymmetry 
measurements, $S_{\pi^+\pi^-}$ and $C_{\pi^+\pi^-}$ and then,  using 
these values, we obtained the constraints  for the color-suppressed 
and color-allowed tree ratio parameters, $C/T$ and $\delta_{CT}$ for different
given values of $\gamma$. We found that the errors in the branching 
ratios are still large and the allowed region for $\delta_{CT}$ is 
distributed in a quite large range. On the other hand, the value of $C/T$ 
is found to be rather large for most of the parameter space and for 
example, we found $C/T \gsim 0.5$ for $\gamma > 47^{\circ}$. 

Next, we examined whether this large value of $C/T$ can be 
explained within the uncertainties of the theoretical model
 computations. We examined two theoretical models, i) QCD factorization 
varying $\rho_{H, A}$ and $\phi_{H, A}$ and ii) QCD factorization 
with $\rho_{H, A}=0$ (no strong phase from perturbative part) but adding 
FSI phase. For i), we found that large $\rho_{H, A}$ lead to large
values of $C/T$, especially when $\delta_{CT}$ is small. On the other 
hand, for ii), we found that $C/T$ and $\delta_{CT}$ are enhanced when 
the FSI phase $\delta_{20}$ increases. As a result, we found that the 
large $C/T$ can be explained in both cases, within the large theoretical 
uncertainties from meson distribution amplitudes, together with 
the end-point singularity for the former and with the FSI phase  for the 
latter. We found that in general, the larger $C/T$ can be realised for 
the smaller $\delta_{CT}$ for case i) and for the larger $\delta_{CT}$ 
for case ii). Therefore we will be able to  distinguish 
these two sources of enhancement factors in near future 
by using the measurement of $C_{00}$. Namely, the ratio to $C_{+-}$ yields 
\be
\frac{C_{00}}{C_{+-}}=\frac{C}{T}\frac{\sin (\delta_{CT}-\delta_{PT})}{\sin \delta_{PT}} \frac{1}{R_{00}}. 
\ee 
One can see that typically, a small $\delta_{CT}(\simeq 0)$ leads to this ratio of order unity with negative sign,  $C_{00}/C_{+-}\simeq -C/T /R_{00}$. For example, the central values of the experimental data
	for $R_{00}$ and $C_{+-}$ lead to $C_{00}=0.57$ for $\delta_{CT}=0$, which is close
	to the higher end of the current experimental value of $C_{00}$ in Eq. (\ref{eq:c00}). 
We can also see that a large  $\delta_{CT}(\simeq \pm \pi/2)$ result shows a strong dependence on $\delta_{PT}$, $C_{00}/C_{+-}\simeq \pm C/T /R_{00}/\tan\delta_{PT}$.
Thus, for a more precise analysis, we will need a better knowledge about $\delta_{PT}$ from measurements of $(S_{+-}, C_{+-})$ as well as the prediction of $\delta_{PT}$ from each model.  
Note that the values of $\delta_{CT}$ and $\delta_{PT}$ are related in
QCD factorisation through the parameters of the end-point singularity
but are independent in FSI, especially due to a possible inelastic re-scattering phase $\delta_D$ of Eq. (\ref{eq:29}). 
 \bigskip

\begin{acknowledgments}
The work by E.K. was supported by the Belgian
Federal Office for Scientific, Technical and Cultural Affairs through the
Interuniversity Attraction Pole P5/27. 
\end{acknowledgments}
 


\begin{thebibliography}{99}
\bibitem{bsbs}
  P.~Ball, S.~Khalil and E.~Kou,
  Phys.\ Rev.\ D {\bf 69} (2004) 115011
  [arXiv:hep-ph/0311361].
\bibitem{Gronau:2002qj}
  M.~Gronau and J.~L.~Rosner,
  Phys.\ Rev.\ D {\bf 65} (2002) 093012
  [arXiv:hep-ph/0202170].
\bibitem{Gronau:2002gj}
  M.~Gronau and J.~L.~Rosner,
  Phys.\ Rev.\ D {\bf 66} (2002) 053003
  [Erratum-ibid.\ D {\bf 66} (2002) 119901]
  [arXiv:hep-ph/0205323].
\bibitem{Ali:2004hb}
  A.~Ali, E.~Lunghi and A.~Y.~Parkhomenko,
  Eur.\ Phys.\ J.\ C {\bf 36} (2004) 183
  [arXiv:hep-ph/0403275].
\bibitem{Charng:2004ed}
  Y.~Y.~Charng and H.~n.~Li,
  Phys.\ Rev.\ D {\bf 71} (2005) 014036
  [arXiv:hep-ph/0410005].
\bibitem{Zenczykowski:2004tw}
  P.~Zenczykowski,
  Phys.\ Lett.\ B {\bf 590} (2004) 63
  [arXiv:hep-ph/0402290].
\bibitem{Chiang:2004nm}
  C.~W.~Chiang, M.~Gronau, J.~L.~Rosner and D.~A.~Suprun,
  Phys.\ Rev.\ D {\bf 70} (2004) 034020
  [arXiv:hep-ph/0404073].
\bibitem{Mishima:2004um}
  S.~Mishima and T.~Yoshikawa,
  Phys.\ Rev.\ D {\bf 70} (2004) 094024
  [arXiv:hep-ph/0408090].
\bibitem{He:2004ck}
  X.~G.~He and B.~H.~J.~McKellar,
  arXiv:hep-ph/0410098.
\bibitem{Gronau:2004sj}
  M.~Gronau, E.~Lunghi and D.~Wyler,
  Phys.\ Lett.\ B {\bf 606} (2005) 95
  [arXiv:hep-ph/0410170].
\bibitem{Buras:2004th}
  A.~J.~Buras, R.~Fleischer, S.~Recksiegel and F.~Schwab,
  Acta Phys.\ Polon.\ B {\bf 36} (2005) 2015
  [arXiv:hep-ph/0410407].
\bibitem{Xing:2005pr}
  Z.~z.~Xing and H.~Zhang,
  Phys.\ Rev.\ D {\bf 71} (2005) 051302
  [arXiv:hep-ph/0501016].
\bibitem{Sowa:2005vn}
  M.~Sowa and P.~Zenczykowski,
  Phys.\ Rev.\ D {\bf 71} (2005) 114017
  [arXiv:hep-ph/0502032].
\bibitem{Grossman:2005jb}
  Y.~Grossman, A.~Hocker, Z.~Ligeti and D.~Pirjol,
  arXiv:hep-ph/0506228.
\bibitem{Imbeault:2005ne}
  M.~Imbeault,
  arXiv:hep-ph/0505254.
\bibitem{Raz:2005hu}
  G.~Raz,
  arXiv:hep-ph/0509125.
\bibitem{Beneke:2000ry}
  M.~Beneke, G.~Buchalla, M.~Neubert and C.~T.~Sachrajda,
  Nucl.\ Phys.\ B {\bf 591} (2000) 313
  [arXiv:hep-ph/0006124].
\bibitem{Beneke:2003zv}
  M.~Beneke and M.~Neubert,
  Nucl.\ Phys.\ B {\bf 675} (2003) 333
  [arXiv:hep-ph/0308039].
\bibitem{melic}
  A.~Khodjamirian, T.~Mannel, M.~Melcher and B.~Melic,
  Phys.\ Rev.\ D {\bf 72} (2005) 094012
  [arXiv:hep-ph/0509049].
  \bibitem{Cheng:2004ru}
  H.~Y.~Cheng, C.~K.~Chua and A.~Soni,
  Phys.\ Rev.\ D {\bf 71} (2005) 014030
  [arXiv:hep-ph/0409317].
\bibitem{Cheng:2005bg}
  H.~Y.~Cheng, C.~K.~Chua and A.~Soni,
  Phys.\ Rev.\ D {\bf 72} (2005) 014006
  [arXiv:hep-ph/0502235].
\bibitem{kim}
  D.~Chang, C.~S.~Chen, H.~Hatanaka and C.~S.~Kim,
  arXiv:hep-ph/0510328.
\bibitem{HFAG}
  H.~F.~A.~Group(HFAG),
  arXiv:hep-ex/0505100.
\bibitem{Buras:2003dj}
  A.~J.~Buras, R.~Fleischer, S.~Recksiegel and F.~Schwab,
  Phys.\ Rev.\ Lett.\  {\bf 92} (2004) 101804
  [arXiv:hep-ph/0312259].
\bibitem{Ball:2003fq}
  P.~Ball and E.~Kou,
  JHEP {\bf 0304} (2003) 029
  [arXiv:hep-ph/0301135].
\bibitem{Braun:2003wx}
  V.~M.~Braun, D.~Y.~Ivanov and G.~P.~Korchemsky,
  Phys.\ Rev.\ D {\bf 69} (2004) 034014
  [arXiv:hep-ph/0309330].
\bibitem{Lee:2005gz}
  S.~J.~Lee and M.~Neubert,
  Phys.\ Rev.\ D {\bf 72} (2005) 094028
  [arXiv:hep-ph/0509350].
\bibitem{Ball:2005tb}
  P.~Ball and R.~Zwicky,
  Phys.\ Lett.\ B {\bf 625} (2005) 225
  [arXiv:hep-ph/0507076].
\bibitem{Bauer:2004tj}
  C.~W.~Bauer, D.~Pirjol, I.~Z.~Rothstein and I.~W.~Stewart,
  Phys.\ Rev.\ D {\bf 70} (2004) 054015
  [arXiv:hep-ph/0401188].
\bibitem{Bauer:2005wb}
  C.~W.~Bauer, D.~Pirjol, I.~Z.~Rothstein and I.~W.~Stewart,
  arXiv:hep-ph/0502094.
\bibitem{Bauer:2005kd}
  C.~W.~Bauer, I.~Z.~Rothstein and I.~W.~Stewart,
  arXiv:hep-ph/0510241.
\bibitem{zupan}
  A.~R.~Williamson and J.~Zupan,
  arXiv:hep-ph/0601214.
  \bibitem{beneke2}
  M.~Beneke and D.~Yang,
  Nucl.\ Phys.\ B {\bf 736} (2006) 34
  [arXiv:hep-ph/0508250]. \\
  M.~Beneke and S.~Jager,
  arXiv:hep-ph/0512351.
\bibitem{Suzuki:1999uc}
  M.~Suzuki and L.~Wolfenstein,
  Phys.\ Rev.\ D {\bf 60} (1999) 074019
  [arXiv:hep-ph/9903477].
\bibitem{Fajfer:2005mf}
  S.~Fajfer, T.~N.~Pham and A.~Prapotnik Brdnik,
  Phys.\ Rev.\ D {\bf 72} (2005) 114001
  [arXiv:hep-ph/0509085].
\bibitem{JM}
  J.~M.~Gerard, J.~Pestieau and J.~Weyers,
  Phys.\ Lett.\ B {\bf 436} (1998) 363
  [arXiv:hep-ph/9803328].
\bibitem{isola}
  C.~Isola and T.~N.~Pham,
  Phys.\ Rev.\ D {\bf 62} (2000) 094002
  [arXiv:hep-ph/9911534].
\bibitem{WW}
  L.~Wolfenstein and F.~Wu,
  Phys.\ Rev.\ D {\bf 72} (2005) 077501
  [arXiv:hep-ph/0506224].
\end{thebibliography}
\end{document}